\documentclass[a4paper]{article}

\usepackage{amsmath}  
\usepackage{amsfonts}
\usepackage{fullpage}
\usepackage{graphicx}
\usepackage{amsmath}
\usepackage{fullpage}
\usepackage{authblk}

\begin{document}
%\begin{frontmatter}

\title{The representation of boundary currents in a finite element shallow water model}

\author{Peter D. D\"uben\thanks{dueben@atm.ox.ac.uk}}
\author{Peter Korn\thanks{peter.korn@mpimet.mpg.de}}
\affil{Max Planck Institute for Meteorology,\\Bundesstrasse 55, 20146 Hamburg, Germany}

%\author[rvt,focal]{Peter Dominik D\"uben\corref{cor1}}
%\ead{dueben@atm.ox.ac.uk}
%\author[rvt]{Peter Korn}
%\ead{peter.korn@mpimet.mpg.de}

%\address[rvt]{Max Planck Institute for Meteorology, Bundesstrasse 55, 20146 Hamburg, Germany}
%\address[focal]{International Max Planck Research School on Earth System Modelling}

\maketitle

\begin{abstract}
We evaluate the influence of local resolution, eddy viscosity, coastline structure, and boundary conditions on the numerical representation of boundary currents in a finite element shallow-water model. The use of finite element discretization methods offers a higher flexibility compared to finite difference and finite volume methods, that are mainly used in previous publications. This is true for the geometry of the coast lines and for the realization of boundary conditions.
For our investigations we simulate steady separation of western boundary currents from idealized and realistic coast lines. The use of grid refinement allows a detailed investigation of boundary separation at reasonable numerical cost.

\end{abstract}

%\begin{keyword}
%\keywords{Finite element; no-slip boundaries; free-slip boundaries; boundary currents; coast line representation \LaTeXe; \emph{\journalabb}}
%\end{keyword}

\section{Introduction}

The mechanisms that lead to separation of boundary currents in the ocean are poorly understood. Numerical models can provide only a small contribution to a better understanding of boundary separation, since the representation of the coast line in today's ocean models differs in essential properties from the coast line in real-world oceans, 
mainly due to the coarse resolution, the high viscosity, and the imprint of the used grid structure. 
This paper investigates boundary separation in finite element models. It is possible that finite element discretization methods will be a common choice to build up future ocean models (see for example the model development in \cite{Danilov2004,Piggott2008,Comblen2009,Dueben2012}). 

For the separation of boundary currents in the ocean, such as the Gulf Stream, there are many possible mechanisms that might influence the position of the separation point such as a change of the direction in the wind field, a potential vorticity crisis, an adverse pressure gradient, boundary conditions, a collision with another western boundary current, interactions with the deep western boundary current, the coast line geometry, the bottom topography or eddy-topography interactions (see for example \cite{Stommel1948,Cessi1987,Chassignet1991,Ezer1992,Haidvogel1992,Oezgoekmen1997,Tansley2000,Munday2005,Chassignet2008} and the references therein).

While the Gulf Stream tends to overshoot the separation point of the real world in standard numerical ocean models, state-of-the-art high-resolution model simulations, with a grid resolution of $1/10^{\circ}$ or higher, obtain an improved representation of Gulf Stream separation \cite{Bryan2007, Chassignet2001}. 
However, high-resolution does not guarantee a proper representation of the Gulf Stream, and the separation point remains sensitive to changes in the model setup, such as changes in viscosity parameterization \cite{Bryan2007}. %There is still not one recipe to obtain a realistic representation of boundary currents and boundary separation in ocean models (\cite{Chassignet2008}).
 The choice of boundary conditions is also known to have a significant influence on the separation behavior \cite{Dengg1993,Haidvogel1992}.

The two discretization methods that are widely used in today's ocean models are the finite difference and the finite volume method. The finite difference method offers only a poor representation of the coast line. To introduce a coast line into a finite difference model, grid points on land are typically removed from a fixed grid. Structured longitude/latitude grids which are typically used allow an angle of 0 or 90 degrees between neighboring grid edges along the boundary. 
This leads to staircase pattern along the coast line. Furthermore, due to the staggering of the velocity components, the effective boundary conditions can be dependent on the angle between the coast line and the coordinate axis of the numerical grid (see \cite{Adcroft1997} for the analysis on an Arakawa C-grid and B-grid). Finite volume and finite element methods offer higher geometric flexibility. In finite volume methods, the velocity field is represented as one dimensional vector perpendicular to grid edges. In finite element methods, the velocity field is typically defined as two-dimensional vector quantity all along the coast line. Both methods allow the use of boundary conforming grid generators, in which the vertices at the boundary of a grid are aligned to the coast line.
Despite the improved coast line representation, a detailed analysis of the properties of boundary currents and boundary separation has not been done for finite element models with realistic coast lines for ocean models.

We study the numerical representation of western boundary currents in a finite element model and compare the results to finite difference simulations from the literature. The used finite element model uses a discontinuous linear representation for velocity and a continuous second order representation for height. It was developed for the particular use in atmosphere and ocean model and fulfills the Ladyzhenskaya-Babuska-Brezzi-condition -- which is a necessary condition for convergence in finite element modeling -- and is able to represent the geostrophic balance at the same time \cite{Cotter2009, Cotter2009LBB}. We simulate the separation of steady western boundary currents from idealized coast lines, and coast lines as used in ocean models. We vary the resolution, the eddy viscosity, the grid structure, the coast line, the alignment between the velocity components and the coast line, and between no-slip and free-slip boundary conditions. 
We evaluate the influence of these properties on boundary currents, and boundary separation. The test cases studied in this publication do not fundamentally differ from test setups used in publications such as \cite{Dengg1993}, \cite{Haidvogel1992}, or \cite{Oezgoekmen1997} for simulations with finite difference models with vorticity as prognostic quantity. The main differences are that we use a finite element model, and velocity and height as prognostic quantities.

In section two, we give a very short description of the model setup, including the shallow-water equations, the discretization in space and time, and grid refinement. In section three, we introduce the test cases and present the numerical results. In section four, we discuss the results.

\section{Model setup}

This section provides a brief introduction to the functionality of the used model, including the shallow-water equations, the discretization in space and time, and the used grids. A detailed description of the model setup can be found in \cite{Dueben2012}.

\subsection{The viscous shallow-water equations}

Our finite element model simulates the viscous shallow-water equations in non-conservative form
\begin{equation}
\label{bo_shallowu}
	\partial_t \mathbf{u} + \mathbf{u} \cdot \nabla \mathbf{u} + f \mathbf{k} \times \mathbf{u} + g \nabla h - \frac{1}{H} \nabla \cdot  \left( H \nu \nabla \mathbf{u} \right) = \frac{\boldsymbol{\tau}^s}{H } - \gamma_f \mathbf{u},
\end{equation}
\begin{equation}
\label{bo_shallowh}
 	\partial_t h + \nabla \cdot \left( H \mathbf{u} \right) = 0, \notag
\end{equation}
where $\mathbf{u}$ is the two dimensional velocity vector, $f$ is the Coriolis parameter, $\mathbf{k}$ is the vertical unit vector, $g$ is the gravitational acceleration, $\nu$ is the eddy viscosity, $\boldsymbol{\tau}^s$ is the surface wind forcing, $\gamma_f$ is the bottom friction coefficient, $h$ is the surface elevation and $H$ is the height of the fluid column given by $H=h-h_b$, where $h_b$ is the bathymetry. The prognostic variables are the surface elevation and the velocity.  %Diffusion is not needed for stability reasons; we incorporated it for the sake of completeness.

The used model can run with either free-slip ($\mathbf{u} \cdot \mathbf{n} = 0 $, and $\partial_\mathbf{n} \mathbf{u} = 0 $ on the boundary $\partial \Omega$), or no-slip boundary conditions ($\mathbf{u} = 0 $ on $\partial \Omega$). 
We apply a weak representation of no-slip boundary conditions of zero tangential velocity. To this end, we add the penalty term $-\sigma \mathbf{u}$ to right hand side of equation \ref{bo_shallowu} for all velocities along the boundary that pushes the tangential velocity along the boundary towards zero. $\sigma$ is a constant that needs to be adjusted experimentally.
All other boundary conditions are realized as strong boundary condition by adjusting the corresponding numerical fluxes through the boundary. 

%We therefore use a weak realization of the condition for zero tangential velocity for no-slip conditions, while all other conditions are realized as strong boundary conditions. %We tested the representation of the weak condition in all model simulations performed, the tangential velocity was always resonably small.

\subsection{Discretization in space and time}

Following the typical finite element approach, we expand the physical fields into sets of basis functions $N_i$ and $M_i$

\begin{equation}
   \mathbf{u} = \sum_{i=1}^{N_u} \mathbf{u}_i N_i \qquad \mbox{ and } \qquad h = \sum_{i=1}^{N_h} h_i M_i.  \notag
\end{equation}

We use a $P_1^{DG}P_2$ finite element approach to discretize the equations. This means that we employ discontinuous linear Lagrange polynomials for the representation of the velocity field ($N_i$), and globally continuous quadratic Lagrange polynomials for the representation of the height field ($M_i$). 
Each triangular cell has three degrees of freedom for each component of velocity located at the vertices of the cells, and six degrees of freedom for the height field located at the vertices and edges. While the degrees of freedom of the height field are shared with the surrounding cells, the degrees of freedom of the velocity field belong to a specific cell, which leads to a discontinuous representation. 

Time integration is performed by an explicit three-level Adams-Bashforth method. The equation

\begin{equation}
  	\partial_t \boldsymbol{\psi} = R(\boldsymbol{\psi}), \notag
\end{equation}
where $R$ denotes the right-hand-side of the system, and $\boldsymbol{\psi}$ is the vector of prognostic variables, is discretized in time by

\begin{equation}
    \boldsymbol{\psi}^{n+1} = \boldsymbol{\psi}^{n} + \Delta t \left( \frac{23}{12} R(\boldsymbol{\psi}^n) - \frac{4}{3} R(\boldsymbol{\psi}^{n-1}) + \frac{5}{12} R(\boldsymbol{\psi}^{n-2}) \right), \notag
\end{equation}
where $\boldsymbol{\psi}^i$ is the vector of state variables at the $i$-th time step.

\subsection{Grids and grid refinement}

We use two types of standard grids on which refinement is performed. The first type of grids are structured triangular grids that provide a uniform coverage of the longitude/latitude space. The grids are derived from rectangular grids by bisecting each rectangular into two triangles. 
The second type of grids are icosahedral geodesic grids that provide a quasi-uniform coverage of the sphere \cite{Baumgardner1985}. We use static h-refinement to refine the interesting area around the coast line. In h-refinement, new grid points are introduced to the grid, to increase the model resolution in regions of specific interest. The influence of grid refinement to the model solution is investigated in \cite{Dueben2013}.

\section{Numerical tests and results}
\label{bo_testcases}

In this section, we present the numerical results for two test cases. We will start with an idealized case that uses straight boundaries, simulating a steady wind driven ocean gyre with a western boundary current that separates at the corner of an obstacle. 
In the second test, we study a more realistic setup to investigate coast lines as used in ocean models simulating a steady wind driven circulation in an Atlantic shaped basin.

\subsection{Ocean gyre with idealized coast lines}
\label{bo_Dengg_TC}

We study an ocean gyre setup on the northern hemisphere. The gyre is driven by a wind forcing in clockwise direction. Due to the change of the Coriolis parameter in the meridional direction, the gyre is intensified towards the western boundary, and a western boundary current develops \cite{Stommel1948,Pedlosky1996}. 
The current is separating from the coast at the edge of a rectangular obstacle. The setup is chosen to be as close as possible to the setup used in \cite{Dengg1993}. Here, Dengg investigated boundary separation in a model that simulates the barotropic vorticity equation.

\begin{figure}[ht!]
   \centering
   \includegraphics[width=0.4 \textwidth, angle=270]{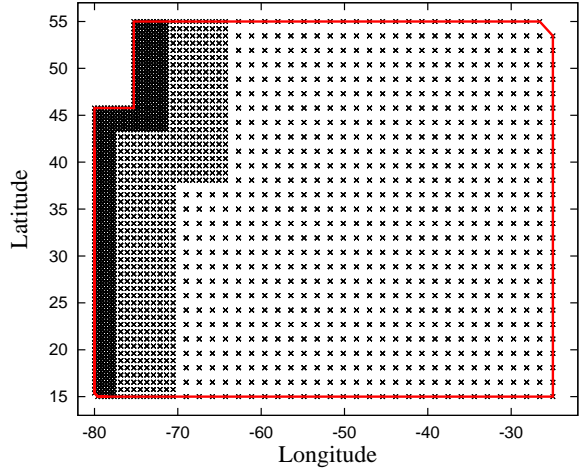}
 \caption{Vertices of the grid with two refinement levels used for the idealized coast line test. The red line marks the coast line.}
  \label{bo_fig:dengg_grid}
  \end{figure}

We perform model runs on a triangular grid, which is structured in longitude/latitude space. We use refinement to increase the resolution in the vicinity of the boundary (see Figure \ref{bo_fig:dengg_grid}). A grid edge has a length of about $1.6^\circ$ in the coarsest and $0.4^\circ$ in the finest part of the grid, this corresponds to about 172 and 43 km at the southern boundary of the domain.

While the meridional wind forcing is zero, the zonal wind forcing is set to

\begin{equation}
 \tau^s_{\lambda} = \tau_0 \cdot 10^{-3} \cdot \cos \left( \frac{\pi \left(\theta - 15^\circ \right)}{40^\circ}  \right) , \notag
\end{equation}

where $\theta$ is the latitude. The bottom friction coefficient $\gamma_f$ is set to $10^{-6} \; s^{-1}$, and $\tau_0$ is $0.28 \; m^2 s^{-2} $. The height field is initialized with a constant water depth of $1000 \; m$; the initial velocity is set to zero.

\begin{figure}[ht!]
   \centering
   \includegraphics[width=0.6 \textwidth, angle=90]{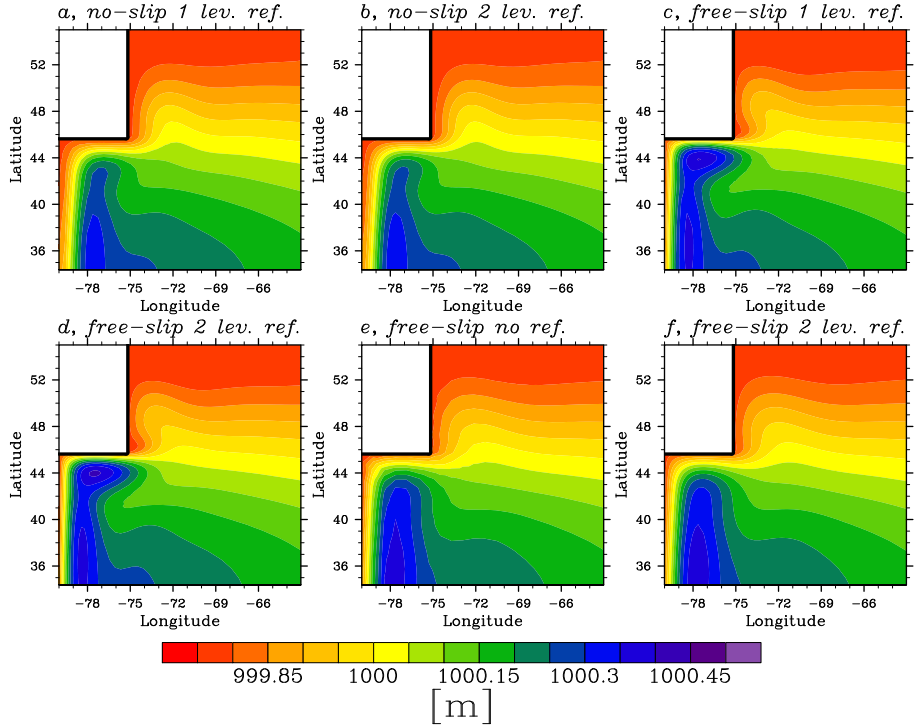}
 \caption{Equilibrium height field of the idealized coast line test. While the runs $\mathit{a}$, $\mathit{b}$, $\mathit{c}$, and $\mathit{d}$ were simulated with $\nu = 3000.0 \; m^2 s^{-1}$, runs $\mathit{e}$ and $\mathit{f}$ were simulated with $\nu = 10000.0 \; m^2 s^{-1}$. 
 The simulations are performed on the grid plotted in Figure \ref{bo_fig:dengg_grid} either without refinement ($\mathit{e}$), with one refinement level ($\mathit{a}$ and $\mathit{c}$), or with two refinement levels ($\mathit{b}$, $\mathit{d}$, and $\mathit{f}$).}
\label{bo_fig:height_resolution}
  \end{figure}

Figure \ref{bo_fig:height_resolution} shows the equilibrated height field. For all tests, the Munk layer at the western boundary is represented smoothly. %The width of the Munk layer is dependent on eddy viscosity. If eddy viscosity is too low, the boundary current is not properly resolved \cite{Griffies2003}. 

The simulations in \cite{Dengg1993} show no boundary separation when free-slip boundary conditions are used. Nevertheless, in the present simulations the boundary flows separate for free-slip and no-slip boundary conditions. The equilibrated fields have a different shape for the two different boundary conditions (compare $\mathit{b}$ and $\mathit{d}$). While resolution does not play an important role for boundary separation (compare $\mathit{a}$ with $\mathit{b}$, $\mathit{c}$ with $\mathit{d}$, and $\mathit{e}$ with $\mathit{f}$), changes in viscosity have a strong impact (compare $\mathit{d}$ and $\mathit{f}$). The simulated flows have Reynoldsnumbers between 10 and 100 along the boundary. The results are qualitatively the same when simulations are performed with stronger wind forcings, and therefore with higher Reynoldsnumbers. We do not show these results since the resulting equilibrium fields are unsteady and it is much more difficult to compare them. 

%\begin{figure}[ht!]
%   \centering
%   \includegraphics[width=0.65 \textwidth, angle=0]{./pictures/Dengg_Bilder/height_bigbox}
% \caption{Equilibrium height field of simulations with idealized coast line and larger obstacle for free-slip (left) and no-slip (right) boundary conditions, with $\nu=3000 \; m^2 s^{-1}$, $\tau_0 = 0.84 \; m^2 s^{-2}$, and `1 level' refinement.}
%\label{bo_fig:vortbb}
%\end{figure}

%Dengg studied a further test in which he simulated a similar test setup as before, with a larger obstacle \cite{Dengg1993}. While he obtained premature separation for no-slip, but not for free-slip boundary conditions. Premature separation means that the flow is not able to follow the northern coast line for a long distance, and separates earlier. Figure \ref{bo_fig:vortbb} shows the equilibrium height field of model runs of the finite element model, performed with a larger obstacle, compared to the previous runs. 
%As in \cite{Dengg1993}, we notice premature separation for no-slip, but not for free-slip boundary conditions. 

%We have performed double gyre simulations similar to the simulations in \cite{Haidvogel1992} (not shown in this paper). These simulations confirm that premature separation takes place for no-slip, but not for free-slip boundary conditions.

\begin{figure}[ht!]
   \centering
   \includegraphics[width=1.0 \textwidth, angle=0]{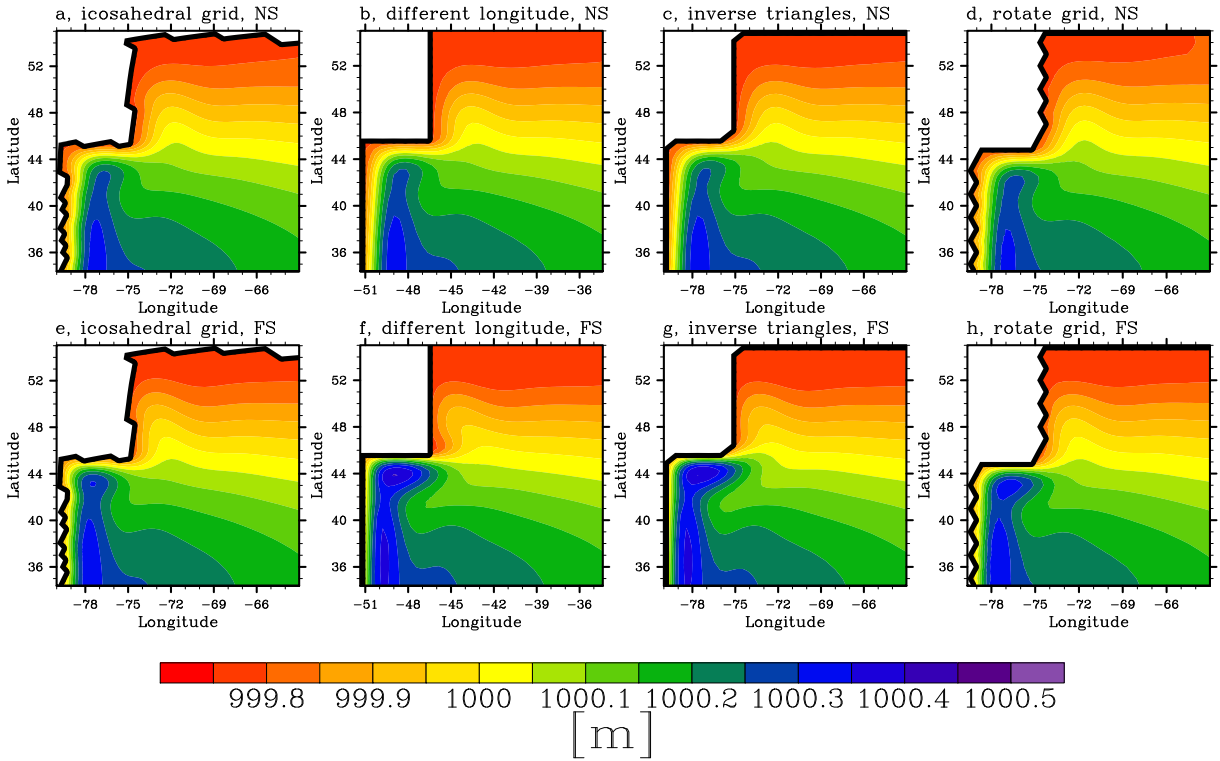}
 \caption{Equilibrium height field of the idealized coast line model runs with no-slip (top row) and free-slip (bottom row) boundaries conditions on different grids. Results should be compared with the model runs $\mathit{a}$ and $\mathit{c}$ in Figure \ref{bo_fig:height_resolution}.}
\label{bo_fig:height_changegrid}
  \end{figure}

\begin{figure}[ht!]
   \centering
   \includegraphics[width=0.4 \textwidth, angle=270]{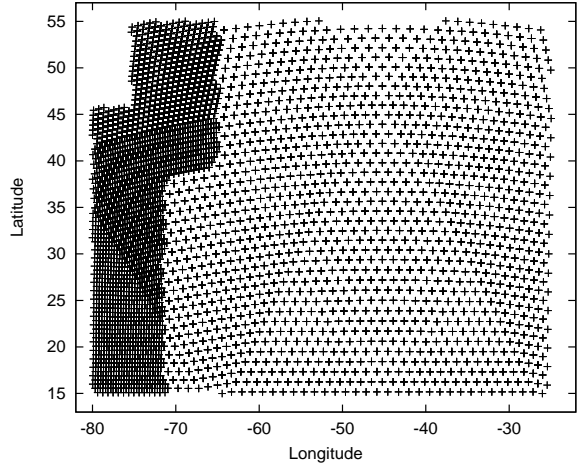}
 \caption{Vertices of the refined grid built from an icosahedral grid, used in model run $\mathsf{a}$ in Figure \ref{bo_fig:height_changegrid}.}
\label{bo_fig:grid_icosahedron}
\end{figure}

\begin{figure}[ht!]
   \centering
   \includegraphics[width=0.2 \textwidth, angle=0]{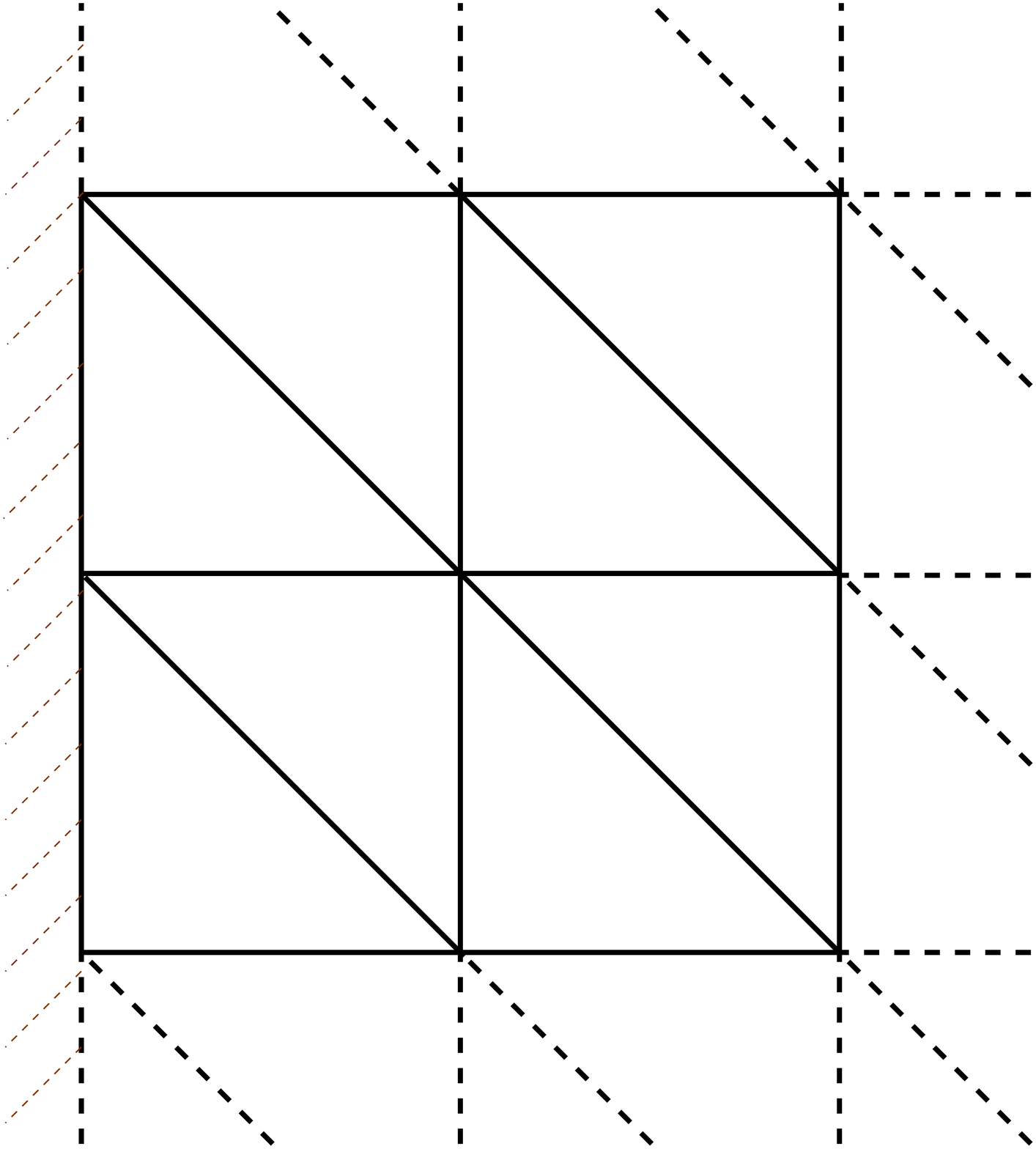} $\qquad$ \includegraphics[width=0.2 \textwidth, angle=0]{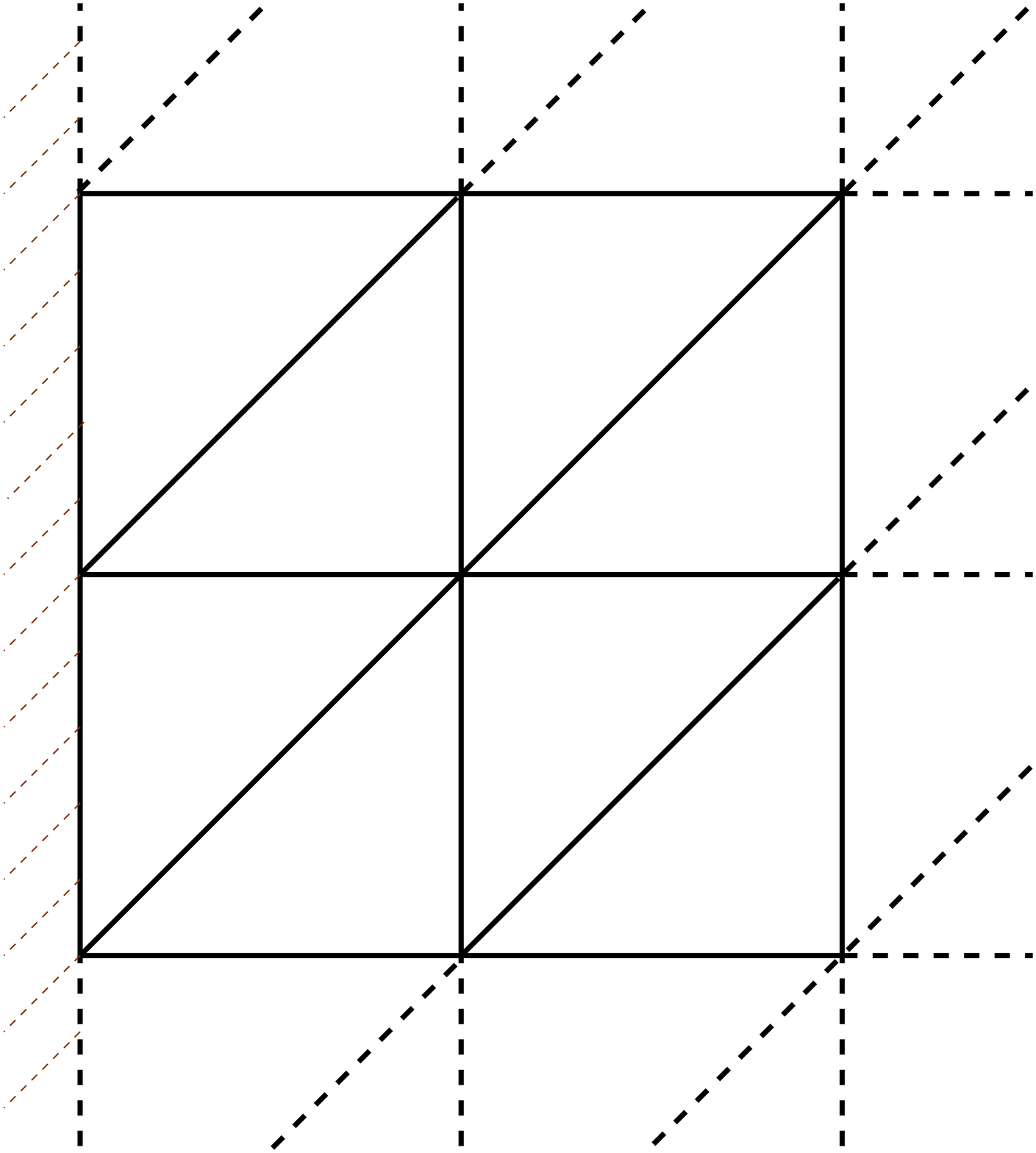} $\qquad$ \includegraphics[width=0.37 \textwidth, angle=0]{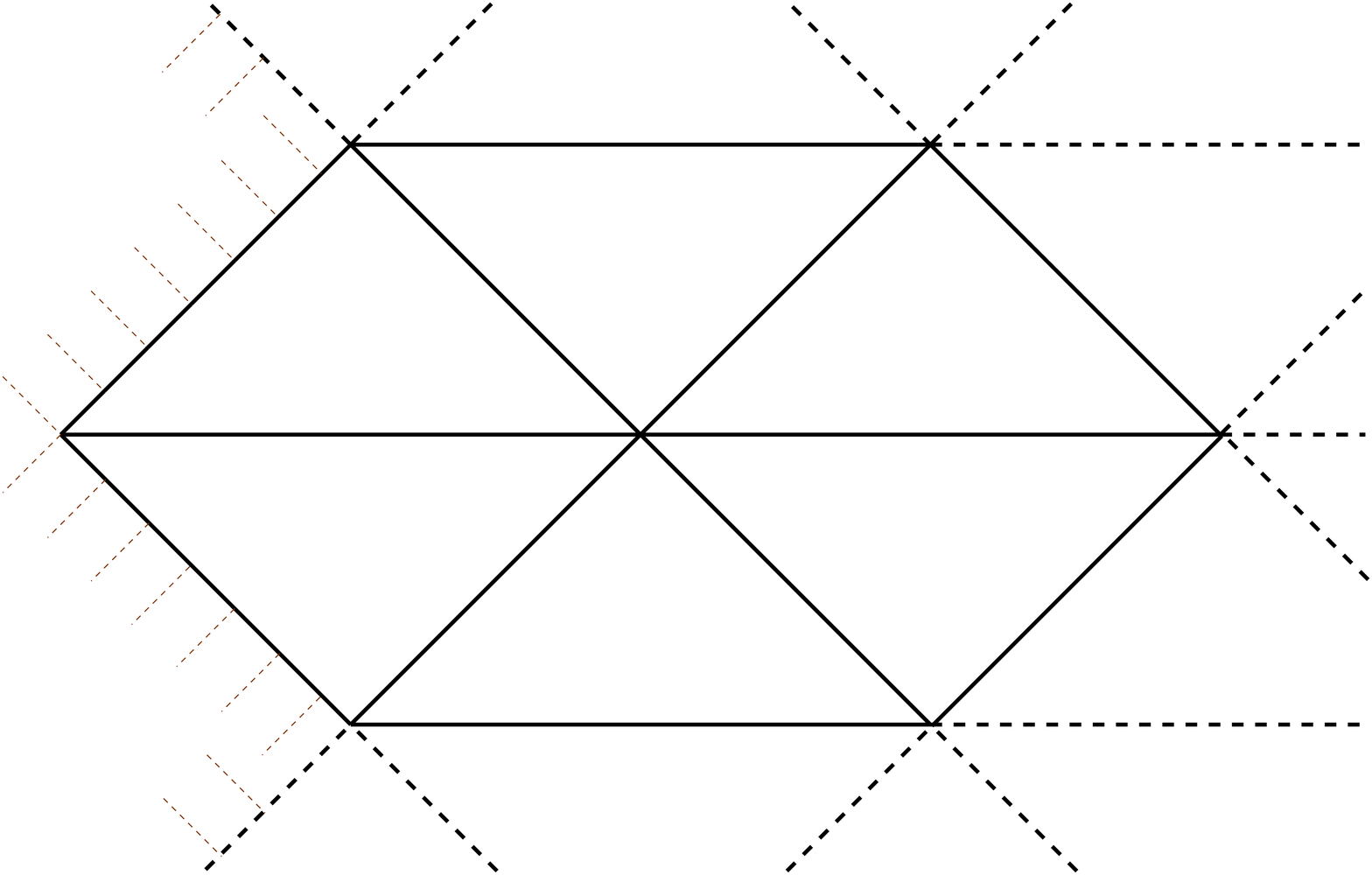}
 \caption{Structure of the triangular grids used for the standard runs (left), simulations $\mathsf{c}$ and $\mathsf{g}$ (middle), and simulations $\mathsf{d}$ and $\mathsf{h}$ in Figure \ref{bo_fig:height_changegrid} (right), with indicated coast line at the western boundary.}
\label{bo_fig:changegrid_grids}
  \end{figure}

The simulations in Figure \ref{bo_fig:height_changegrid} are performed with the same setup as for the previous idealized coast line tests $\mathit{a}$ and $\mathit{c}$ in Figure \ref{bo_fig:height_resolution} with $\nu = 3000 \; m^2 s^{-1}$, $\tau_0 = 0.28 m^2 s^{-2} $. However, different grids have been used. Simulations $\mathsf{a}$ and $\mathsf{d}$ were performed on a one level refined icosahedral grid (see Figure \ref{bo_fig:grid_icosahedron}). The coast line is represented by an unstructured pattern. 
The simulations $\mathsf{b}$ and $\mathsf{f}$ were performed on a grid with a changed longitude of the model domain compared to the standard grid. A change of the longitude is changing the alignment of the two components of the velocity field $u$ and $v$ with the coast line. In the model runs $\mathsf{c}$, $\mathsf{d}$, $\mathsf{g}$ and $\mathsf{h}$ the arrangement of the triangles in the structured grid was changed (Figure \ref{bo_fig:changegrid_grids}). The change of the grid structure leads to a zig-zag pattern of the coast line in model run $\mathsf{d}$ and $\mathsf{h}$.

While all simulations with no-slip boundary conditions result in a fairly similar gyre structure, this is different for the free-slip simulations. The simulations with free-slip boundary conditions and unstructured or zig-zag meridional coast line ($\mathsf{e}$ and $\mathsf{h}$) appear to be similar to the no-slip simulations. 
A likely reason for this similarity is a shift of the boundary flow into the interior of the domain that is caused by the boundary conditions in the no-slip case and by the abrupt changes of the direction of the coast line in simulations $\mathsf{e}$ and $\mathsf{h}$.
A similar behavior can be found for simulations with ocean models based on an Arakawa C-grid finite difference scheme that change the boundary conditions from free-slip to no-slip when zig-zag coast lines are considered (see \cite{Adcroft1997}), but the mechanism is different. In the finite element model, the representation of the boundary conditions stays the same when the alignment of the velocity components with the boundary change. This is confirmed by the simulations $\mathsf{b}$ and $\mathsf{f}$ that show almost identical flow pattern compared to simulations $\mathit{a}$ and $\mathit{c}$ in Figure \ref{bo_fig:height_resolution}.

While the separation behavior in the free-slip simulation $\mathsf{g}$ in Figure \ref{bo_fig:height_changegrid} shows differences to the reference simulation $\mathit{a}$ in Figure \ref{bo_fig:height_resolution}, the separation behavior of the no-slip simulation $\mathsf{c}$ is very similar to the reference simulation. The slight change of the coast line is able to change the solution of the free-slip simulation, while this is not true for the no-slip simulation.

\subsection{Irregular coast lines - An Atlantic shaped ocean domain}
\label{bo_Atlantic_TC}

\begin{figure}[ht!]
   \center
   \includegraphics[width=0.36 \textwidth, angle=270]{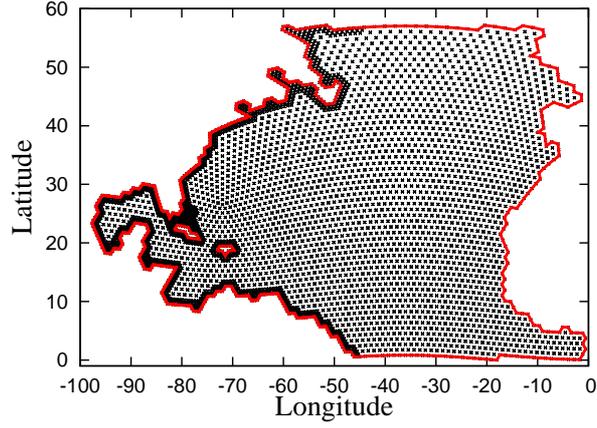}
 \caption{Vertices of the refined grid used for the Atlantic test. The red line marks the coast line.}
  \label{bo_fig:atlantic_grid}
\end{figure}

In this section, we investigate a more realistic application to model the ocean. We simulate an ocean basin which is shaped like the Atlantic ocean and offers a realistic representation of the coast line. The domain is cut at the equator and at 58$^{\circ}$ North. We simulate on real-world topography, but water depth is cut at 1000 m. An artificial wind forcing that is balanced by bottom friction induces a steady circulation. The used numerical grid is plotted in Figure \ref{bo_fig:atlantic_grid}. The grid is refined at the western boundary and has a typical edge length of 120 km in the coarse, and 60 km in the fine part of the grid. In the refined area along the coast line there are always two neighbored grid edges that are aligned with each other. 

Simulations are initialized with zero surface elevation and zero velocity. The zonal wind forcing is given by
\begin{align}
 \tau^s_{\lambda} = \begin{cases}
  -\tau_0 \cdot 10^{-3} \cdot \cos \left( 4 \cdot \theta  \right) \qquad \qquad&\text{if} \qquad \theta < 45^{\circ} \\ 
  0  \qquad &\text{if} \qquad  \theta \ge 45^{\circ}, \notag
\end{cases}
\end{align}

the meridional wind forcing is zero. The bottom friction coefficient $\gamma_f$ is set to $10^{-6} \; s^{-1}$, and $\tau_0$ is $3.0 \; m^2 s^{-2} $.

\begin{figure}[ht!]
   \center
   \includegraphics[width=0.9 \textwidth, angle=90]{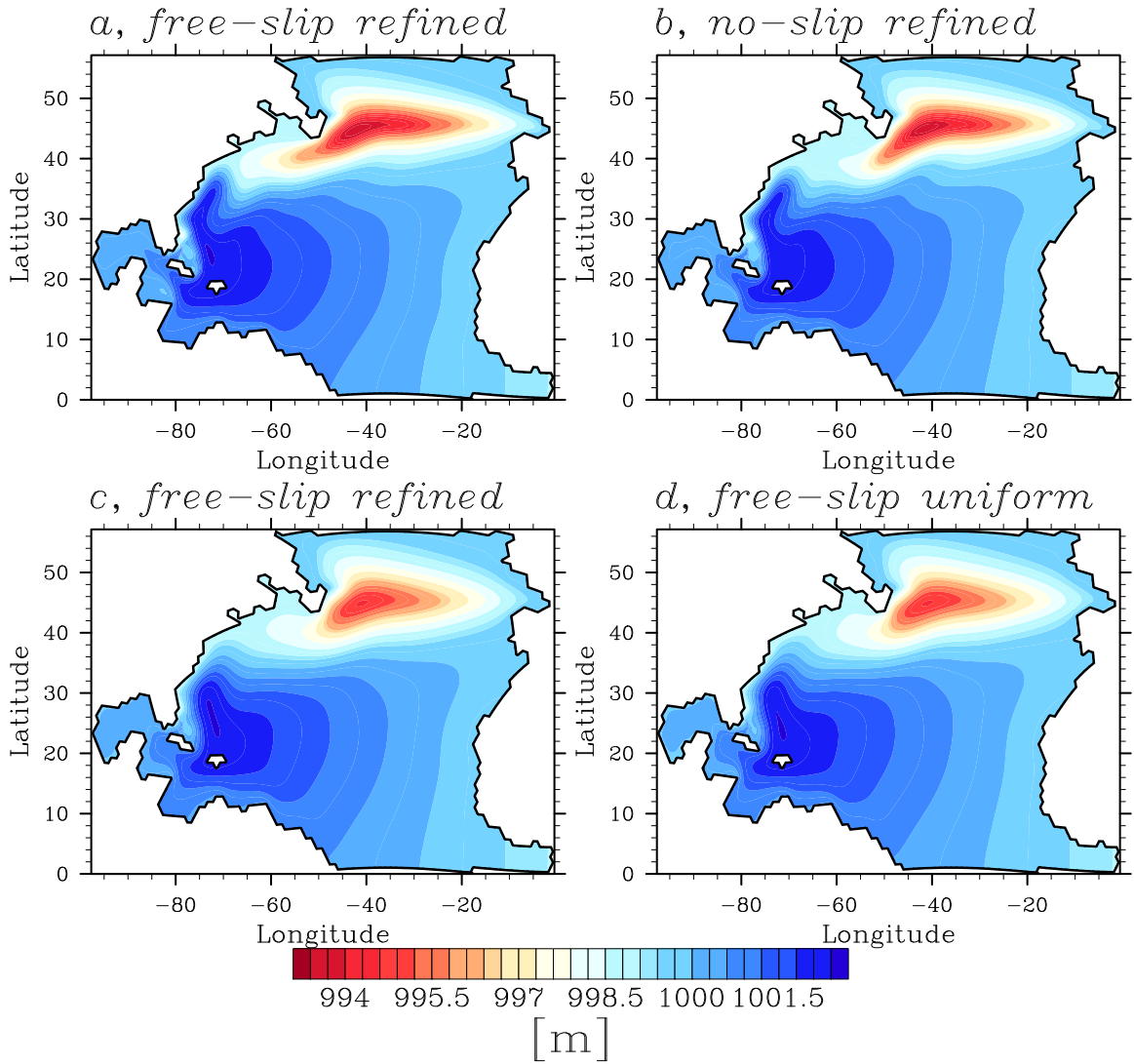} 
 \caption{Equilibrium height field in the Atlantic model runs, after 140 days. While run $\mathit{a}$ and $\mathit{b}$ were simulated with $\nu = 6655.0 \; m^2 s^{-1}$, run $\mathit{c}$ and $\mathit{d}$ were simulated with $\nu = 53240.0 \; m^2 s^{-1}$.	}
  \label{bo_fig:atlantik}
\end{figure}

Figure \ref{bo_fig:atlantik} shows the equilibrium height field of the performed model runs. %As mentioned in the previous subsection, the eddy viscosity fixes the width of the Munk layer \cite{Griffies2003}. A viscosity significantly lower than $6655 \; m^2 s^{-1}$ for the refined, and $53240 \; m^2 s^{-1}$ for the unrefined model run, would lead to an insufficiently resolved Munk layer, which would trigger model instabilities. 

The model runs $\mathit{a}$, $\mathit{b}$ and $\mathit{c}$ are performed on the grid plotted in Figure \ref{bo_fig:atlantic_grid}. Model run $\mathit{d}$ is using the same grid without refinement of the western boundary. As for the simulation with an unstructured representation of the idealized coast line ($\mathsf{a}$ in Figure \ref{bo_fig:height_changegrid}) the unstructured realistic coast line seems to make the used boundary condition play only a minor role -- the height field along the western boundary differs much more for different values of eddy viscosity (compare $\mathit{a}$ with $\mathit{c}$) than for no-slip and free-slip boundary conditions (compare $\mathit{a}$ and $\mathit{b}$). A change in resolution leads to minor changes (compare $\mathit{c}$ and $\mathit{d}$), although it limits the smallest possible value for eddy viscosity (see also \cite{Dueben2013}).

\section{Discussion of the results}

Although finite element methods provide an improved coast line representation compared to finite difference methods, our investigations show that the representation of the coast line and the boundary conditions is still not satisfying. Changes of the grid structure can lead to changes of the separation behavior (see Figure \ref{bo_fig:height_changegrid}).

Our tests on the influence of resolution and eddy viscosity show that steady western boundary currents are not strongly affected by changes in resolution, as long as the Munk layer is resolved properly. However, a higher resolution allows the use of a smaller eddy viscosity, which can change the model results significantly. To this end, grid refinement can be used to increase the local resolution in the Munk layer. 

The model results change strongly between free-slip and no-slip boundary conditions, when idealized straight coast lines are simulated (subsection \ref{bo_Dengg_TC}). This is heavily discussed in the literature (see \cite{Dengg1993} as one example). On the other hand, the model results change only slightly with the boundary conditions for coast lines as used in ocean models (subsection \ref{bo_Atlantic_TC}).
In simulations with free-slip boundaries and zig-zag or unstructured coast lines, the flow is shifted towards the interior of the domain, due to the rapid changes of the direction of the coast line. The results look similar to no-slip model runs, in which the flow is shifted into the interior of the domain via the boundary conditions (subsection \ref{bo_Dengg_TC}).

In contrast to finite difference models with the vorticity as prognostic quantity \cite{Dengg1993}, we obtain separation for free-slip boundary conditions, using a finite element model with velocity and height as prognostic quantities. We do obtain premature separation for no-slip, but not for free-slip boundary conditions (subsection \ref{bo_Dengg_TC}). This result is consistent with results of finite difference models.

Although finite element methods offer an improved coast line representation compared to finite difference methods, the representation of boundary flows remains dependent on the pattern of the coast line, which is -- for today's ocean models -- very much dependent on the resolution, and not satisfying. Small changes of the grid structure can lead to changes in the separation behavior.

\subsection*{Acknowledgments}

We thank David Marshall for a useful revision of a previous version of this paper.

\bibliographystyle{alpha}
\bibliography{./paper}

\end{document}